\documentclass[conference]{IEEEtran}
\IEEEoverridecommandlockouts
\usepackage{amsmath,amssymb,amsfonts}
\usepackage{algorithmic}
\usepackage{graphicx}
\usepackage{textcomp}
\usepackage{xcolor}
\usepackage{comment}

\usepackage[backend=biber,style=numeric-comp,doi=false,url=true,eprint=false,sorting=none,maxbibnames=40]{biblatex}
\usepackage[dvipsnames,svgnames,table]{xcolor}
\usepackage[
    colorlinks=true,
    urlcolor=RoyalBlue,
    linkcolor=ForestGreen,
    citecolor=ForestGreen
]{hyperref}
\DeclareSourcemap{
  \maps[datatype=bibtex]{
    \map{
      \step[fieldset=abstract, null]
    }
  }
}
\AtBeginBibliography{%
  \footnotesize
  \setlength{\bibitemsep}{0pt}
  \setlength{\itemsep}{1pt}
}

\begin{document}

\title{A Compression Tree Generation Method for \\
 High Performance VLSI Datapaths
}

\author{
 \IEEEauthorblockN{Christophe Giacomotto}
 \IEEEauthorblockA{\emph{Ampere Computing, San Jose, California, USA}}
 \IEEEauthorblockA{cgiacomotto@gmail.com}
}

\maketitle

\begin{abstract}
Compression trees have been historically a key component of several arithmetic units, most notably multipliers. The latter have been for the most part standalone units and in common situations synthesis tools can generate very close to optimal solutions. With large matrix multiply structures taking more of a central role in the compute industry, the need for a better understanding of these compression trees is apparent. Furthermore, the boundary of its usage needs to be redefined: dot products, polynomial fitting of non-linear operators, are all sums of products… And since products are sums of partial products, in the end, we are just adding many things together as bit heaps; i.e. the fundamental role of a compression tree. Therefore, we must see them as more than just a component of a multiplier. 
This paper presents a delay-aware compression-tree methodology based on a weighted bit heap. Each bit is described by its significance and relative arrival time, and the generator uses technology-independent normalized delay units to estimate Boolean logical pressure. By exploiting the commutativity of addition within each column, the method reorders bits across reduction stages and uses a Dadda-style maximum-bandwidth topology with flexible CSA3:2 operators. For Radix-4 Booth multipliers from 11 to 64 bits, the resulting trees show lower modeled delay and estimated logic volume than the evaluated Wallace-style maximum-compression trees. Fused four-way integer dot products further demonstrate how the same representation can evaluate the cost of microarchitectural constraints, such as preserving individual product redundant formats. These results show that weighted bit-heap generation provides a practical technology-independent basis for exploring compression trees beyond standalone multiplication.
\end{abstract}


\section{Introduction}
The advent of machine learning algorithms has dramatically shifted the landscape of arithmetic needs. Even though these models, number formats, and operators evolve rapidly, they remain at heart statistical representations requiring many multiplications and additions. However, because of this instability, we see a range of micro-architectural solutions such as large systolic arrays \cite{goog17, apple26, gaudi24}, outer-product array \cite{Nvidia18}, wider vector engines \cite{goog17} and new fused dot product instructions\cite{arm26}. That growth in MAC density naturally pushes datapath designs into fused sums of products with new techniques to improve efficiency in VLSI \cite{lutz24}. Additionally, because of this uncertainty, fast turnaround and application specific solutions with ASICs and FPGAs are appealing. Evidently, they are confronted to the same ``many-addition'' problem but with completely different criteria dictated by these platform limitations, with many recent advances in tree generation topologies \cite{dedinechin24}.

Compression trees are traditionally presented as parts of standalone multipliers. For that use case, the partial-product pattern is known in advance and synthesis or a specialized multiplier generator can select an effective topology. More general datapaths are less regular. Partial products from several operations may be fused, intermediate redundant formats may be shared, and inputs may arrive at different times. Microarchitectural requirements may also require individual products or intermediate redundant formats to remain available. These constraints change the weighted bit heap and can make a tree that is effective for an isolated multiplier unsuitable for the complete datapath.

This paper develops a generator that treats the compression tree as an independent reduction structure. Its input is a weighted bit heap: each relevant bit is specified by its binary position and relative arrival time, without requiring the generator to know whether the bits originated from a multiplier, a dot product, or another arithmetic operation. The methodology uses a technology-independent delay model based on normalized boolean logical pressure, tracks the carry paths of CSA3:2 operators, and reorders commutative bits to balance delay across the tree. This approach closely follows the contributions from \cite{Vojin96, galal13}. Its central topology principle is a Dadda-style maximum bandwidth between reduction stages, which allows early-arriving bits in less-populated columns to be used in later stages.

The contributions of this work are:

\begin{itemize}
\item A weighted bit-heap representation that captures bit significance and relative arrival time independently of the originating arithmetic operation;
\item A flexible, CSA3:2-based generation methodology that uses delay-aware bit ordering and a Dadda-style maximum-bandwidth topology; and
\item An evaluation on Radix-4 Booth multipliers and fused four-way integer dot products that compares modeled delay and estimated logic volume while quantifying the cost of preserving intermediate redundant formats.
\end{itemize}

The resulting model is intended for early design-space exploration and microarchitectural comparison. It is not a replacement for technology-specific cell characterization, synthesis, placement, and routing; rather, it provides a compact way to compare candidate compression structures before those stages.

\section{Logical Effort}

The idea of using Logical Effort(LE) \cite{le91}, or more simply using some form of normalized delay per logical stage as a criterion for choosing summation grouping has been proven to be effective in several studies \cite{Vojin96, nedovic01, townsend03, galal13}. Production digital VLSI designs nowadays typically use standard cell methodologies for datapaths. Larger unconventional custom cells rarely provide an advantage against synthesis tools besides dense memory structures. Similarly, forced cell relative placement and constrained connectivity tend to create severe external placement and routing congestion issues which are far more costly than the marginal local gains. So, it is important to think of gates as logical or boolean pressure stages rather than physical interconnecting cells. Once a boolean equation is set to be synthesized, placed, and routed; the physical chain of standard cells will vary multiple times along that process. For example, while fine tuning the delay, Design Automation (EDA) tools may use unexpected complex gate and refactor the boolean equations differently simply because that particular logic is placed or routed advantageously compared to another similar set of logic. The trend of better transistor improvement compared to wire RC performance per technology node iteration has been a long standing \cite{wires01} and a worsening \cite{wires22} problem which has exacerbated this effect. However, this process is bound and limited by formal logical equivalence with limited scope. In the case of compression trees, even though the bits in a column are mathematically commutative, modifying the order changes the boolean equations of the resulting redundant format (RDF). Evidently, after the Carry Propagating Adder (CPA), the equivalence would be satisfied but unfortunately that can be out of reach - in logical depth - for synthesis tools formal equivalence engines, especially for non-standard structures. Therefore, it is critical to correctly select the grouping based on the aggregate logical pressure of each bit as part of the micro-architectural specifications.

Calculating the correct effort according to LE is meant for fast static timing estimation and is dependent on the technology node used, which is counterproductive for this work. We only need to confirm that the boolean expressions of \eqref{eq1} and \eqref{eq2} typically yield similar delays in dense low-fanout datapath situations, which is still true in most modern standard cell based CMOS nodes \cite{Vojin96, galal13}.

\begin{equation}
X \oplus Y\label{eq1}
\end{equation}
\begin{equation}
X \cdot Y + X \cdot Z + Y \cdot Z\label{eq2}
\end{equation}

\begin{figure}[htbp]
\centerline{\includegraphics{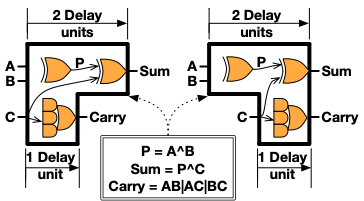}}
\caption{Delay unit normalization and CSA3:2 specifications.}
\label{fig01}
\end{figure}

Under this assumption, we can model the traditional CSA3:2 operator as the boolean equations shown in Fig.~\ref{fig01}. The concept of ``Delay units'' in this context has nothing to do with standard cell delay characterization or even physical breakdown of stages. The term simply refers to ``boolean logical term pressure/effort normalized to equation~\eqref{eq1} or~\eqref{eq2} which are assumed to be equivalent for this normalization''. Additionally, with regard to Fig.~\ref{fig01}, it is important to note that the one ``unit of delay'' for the path from ``C'' to ``Carry'' is floating relative to the ``Sum'' arrival time when its 3-way XOR is specified as 2 stages. Correctly interconnecting the CSA3:2 operators, when specified as in Fig.~\ref{fig01}, can achieve the same delay performance as other larger operators \cite{galal13}. These large operators, like the traditional CSA4:2 for example, are convenient to guarantee the minimum logical depth, i.e. 3 delay units. However, since we are interested in the optimal boolean expression for each final RDF bit, constraining a path with a forced combination limits possibilities. Of course, keeping track of the ``C'' input while figuring out the compression tree groupings is paramount to similarly guarantee minimum logical depth. 

\section{Tree Topologies}
As EDA tools taught us to not be so attached to cells and essentially view compute datapath design as boolean equations instead of chains of standard cells, getting back to the primordial 3:2 counter to provide maximum flexibility helps but is still insufficient. We must also look again at the original tree topologies based on those basic counters. Yet, regardless of the topology, the number \emph{N} of serially dependent stages is dictated by the reduction capability of the 3:2 counter, the tallest part of the tree height ``\emph{H\textsubscript{T}}'', and the shape of the bit heap. Equation~\eqref{eq3} can give an estimate of the number of stages N. But for narrow bit width of \emph{H\textsubscript{T}}, the carries column offset can prevent the last iteration.

  \begin{equation}
  \begin{array}{c@{\qquad}c}
  H_{i+1} = \left\lceil \dfrac{2H_i}{3} \right\rceil
  &
  \begin{gathered}
  H_0 = H_T,\\[2pt]
  N = \min \left\{ i \in \mathbb{Z}_{\geq 0}
  \;\middle|\; H_i \leq 2 \right\}
  \end{gathered}
  \end{array}
  \label{eq3}
  \end{equation}

The philosophies behind Wallace \cite{wallace64} and Dadda \cite{dadda65} trees are
straightforward:

\begin{itemize}
\item
  A Wallace tree will reduce the maximum amount of bits in parallel at
  every stage
\item
  A Dadda tree will reduce the minimum required amount of bits in
  parallel at every stage
\end{itemize}

In essence, Dadda retains maximum bandwidth between stages and Wallace minimizes it. That means these two approaches are not arbitrary restrictive algorithms, they simply represent the extreme bounds of the bit heap bandwidth retention management between reduction stages. The fundamental issue with minimizing the bandwidth at each column is the overuse of the CSA2:2 as it does not reduce that bandwidth and is hence an inefficient operator. Limiting the use of CSA2:2 increases the stage to stage bandwidth and combining this with CSA4:2s typically gives decent overall delay and area results \cite{bickerstaff93}. When a manual implementation is microarchitecturally needed, these hybrid trees have been the preferred solution in a production environment. Unfortunately, they could be faster and smaller. We show in this paper that allowing flexibility with the exclusive use of the base CSA3:2 counter as well as the maximum bandwidth retention from the Dadda tree is best under the proposed delay model.

However, under certain conditions, we may need to reduce the stage-to-stage bandwidth anyway. For example, if the tree needs to be pipelined due to tight frequency requirements or even too much glitching \cite{larsson25}, it may be preferable to limit the number of bits stored in the pipeline registers since they can contribute significantly to the cost in area and power. That being said, nowadays most compression trees fit in a cycle - even for 64b multipliers - which wasn't the case in the past: State elements growth factor in pipelined trees used to force Wallace tree topologies to limit power consumption \cite{zyuban04}. So, previously, the real question was: How much bandwidth between stages a particular tree implementation can afford? That alone made the Dadda tree structure unusable in those situations.
\section{Generator Methodology}
Arguably, the most important part of a generator is its ease of use and versatility. Historically, generators have been built to study how a particular operation or function scales \cite{galal13} which is great to understand specific arithmetic operator design nuances.
However, in a production environment we can't use them: Datapaths are shared, internal settings for multiple micro-instructions support and custom system configurations make the bulk of the complexity in real world designs. Furthermore, as a result of these forced functional divergences, repipelining or timing optimizations would alter how the tree would have been generated for a unique function. Hence, we need a generator in its most simplistic form: the tree alone. 

\begin{figure}[htbp]
\centerline{\includegraphics[scale=0.25]{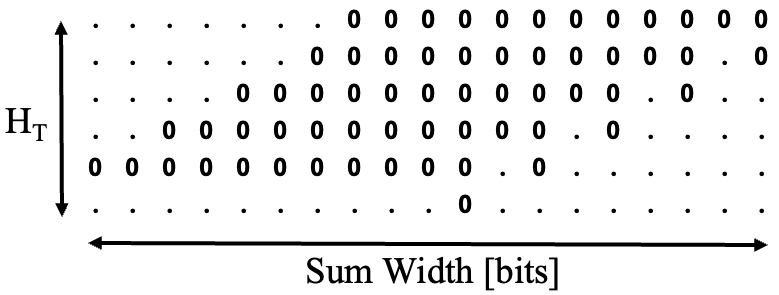}}
\caption{Generator source example, Bit heap specification}
\label{fig02}
\end{figure}

As an example, we will use the familiar partial products of a standard 9x9 Radix-4 Booth multiplier \cite{booth51}. The easiest way to represent multiple binary numbers of different widths and alignment -to be added together- is to write them as a rectangular matrix of characters as in Fig.~\ref{fig02}. We will use ``.'' for the bit positions to be ignored and an integer value for the bit positions to be compressed. The value represents the relative arrival time of that bit compared to the other bits in units of delay as expressed in Fig.~\ref{fig01}. In this example, all the bits to be compressed arrive at the same time, hence they are labeled as ``0'': The earliest arrival time.

In essence, Fig.~\ref{fig02} describes the weighted bit heap \cite{dedinechin24} input specification to be compressed. A compression tree does not need to know more information than this to be generated as a logically balanced structure. In fact, it does not even need to know we are multiplying -or any other arithmetic construct for that matter- aside from: ``we are adding all these things together and that is when the relevant bits arrive''. For this example, a quick simple heuristic would be to note the tallest column is 7 bits, that would mean a CSA4:2 and a CSA3:2 in parallel and then a dependent CSA4:2 to get the final RDF. Assuming 3 units of delay for a CSA4:2, we should then expect a total of 6 units of delay for the worst path of that compression tree.

\begin{figure}[htbp]
\centerline{\includegraphics[width=1\linewidth]{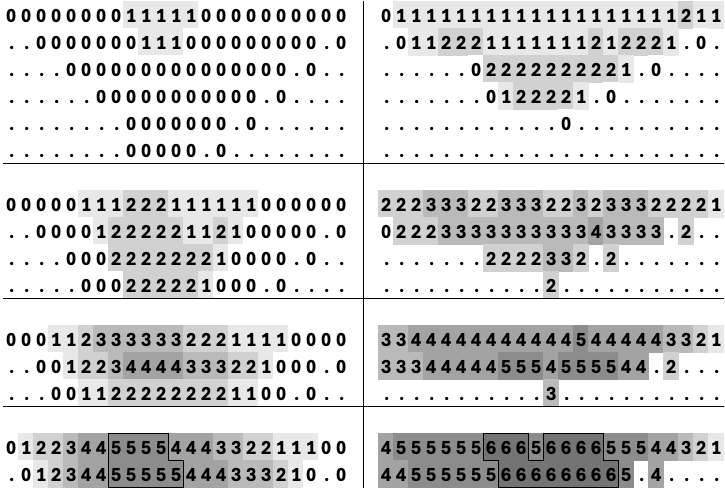}}
\centerline{(a)\hspace{120pt}(b)}
\caption{Example comparison of delay results from generated trees}
\label{fig03}
\end{figure}

The construction of the tree is shown as the bit heap status between
stages in Fig.~\ref{fig03}. Each stage describes the
weighted bit heap output and subsequently input specification to be
compressed for the next stage. If we generate this tree using the
proposed delay balancing methodology with maximum compression at every
stage, we do get a delay of 6 as shown in Fig.~\ref{fig03}(b). However, when allowing maximum bandwidth between stages, we can
see in Fig.~\ref{fig03}(a) that previously uncompressed
bits with early arrival time can opportunistically feed in the high
logical pressure inputs of the CSA3:2 in later stages. This mechanism
contributes to an overall faster delay on the sides of the tree where
column height is less than the maximum. These faster non-critical paths
directly contribute to less logical pressure on the center critical
paths, allowing a better outcome.

\begin{figure}[htbp]
\centerline{\includegraphics[width=1\linewidth]{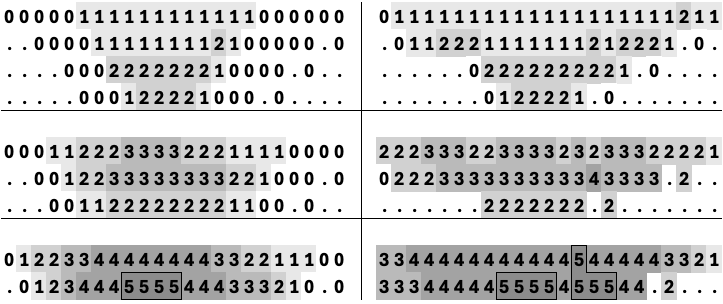}}
\centerline{(a)\hspace{120pt}(b)}
\caption{Example comparison of delay without the 7\textsuperscript{th} partial product}
\label{fig04}
\end{figure}

It can be argued that for this specific example, the 7th partial product is really just a single sign correction bit and it can potentially be ignored or refactored later/differently under hypothetical conditions. By following the same heuristic as before, a height of now 6 bits means two CSA3:2 in parallel and a CSA4:2 to finish, yielding an expected total delay of 5 units. And yet, still, the resulting compression tree with maximum bandwidth retention (Fig.~\ref{fig04}a) ends up overall with faster and/or less critical paths than a Wallace tree (Fig.~\ref{fig04}b) even if the worst delay matches up at 5 units this time. More importantly, it shows the Dadda tree was capable in this case to limit the impact of the delay cost of that single extra bit even though it initially needed to be added to the tallest section of the tree.

\begin{figure}[htbp]
\centerline{\includegraphics[width=1\linewidth]{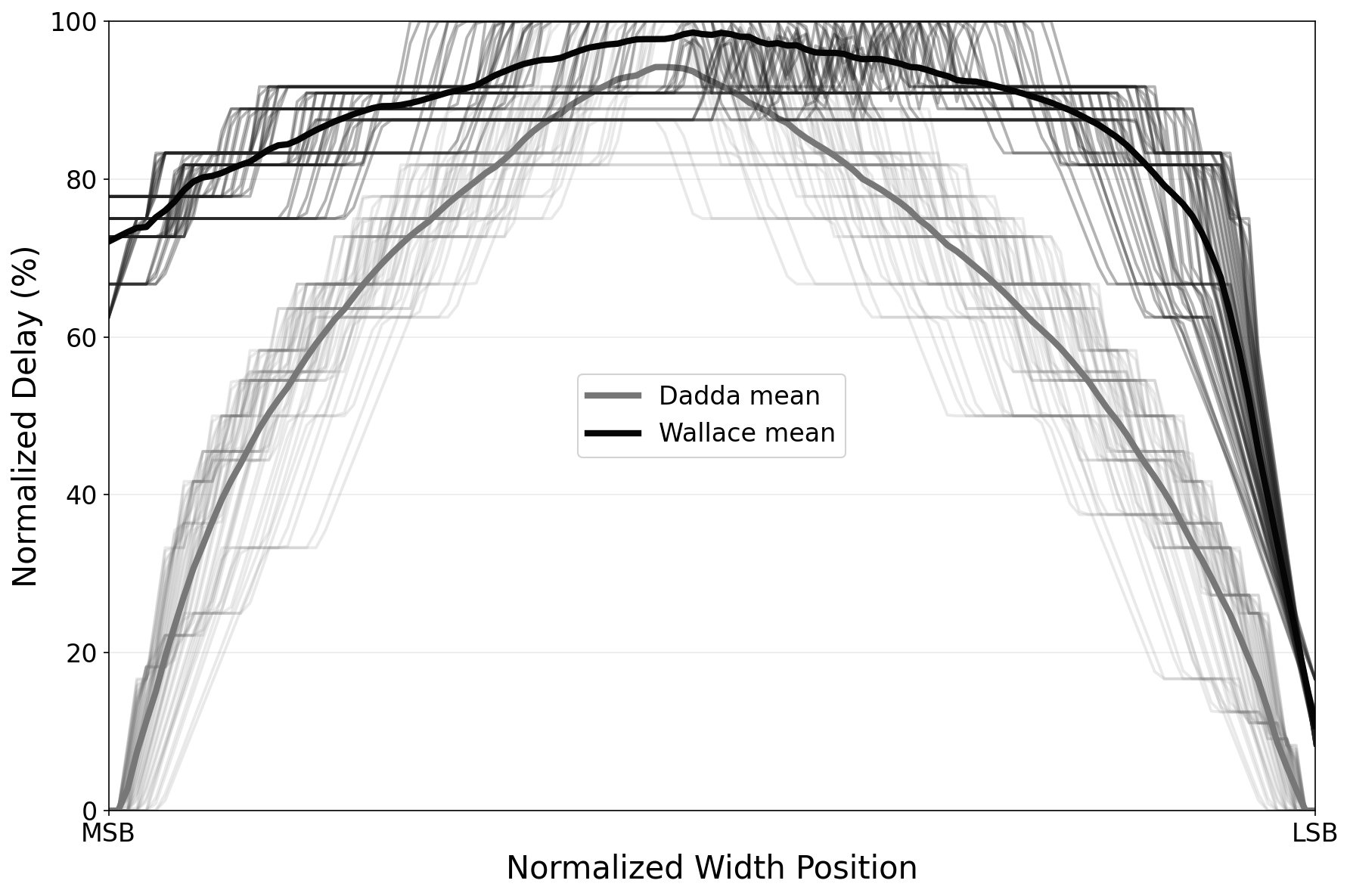}}
\caption{Normalized delay comparison per bit position of every Radix-4 Booth multiplier from 11b to 64b}
\label{fig05}
\end{figure}

As we go beyond that one example and generate the compression trees for every Radix-4 Booth multiplier up to 64b in Fig.~\ref{fig05}, we can then see the shape of the delay distribution stays consistent for both tree topologies. Interestingly, the Dadda tree remains systematically better irrespective of the multiplier size. The approximately smooth, symmetric delay profile is consistent with the shape of the underlying partial-product bit heap.

\begin{figure}[htbp]
\centerline{\includegraphics[width=1\linewidth]{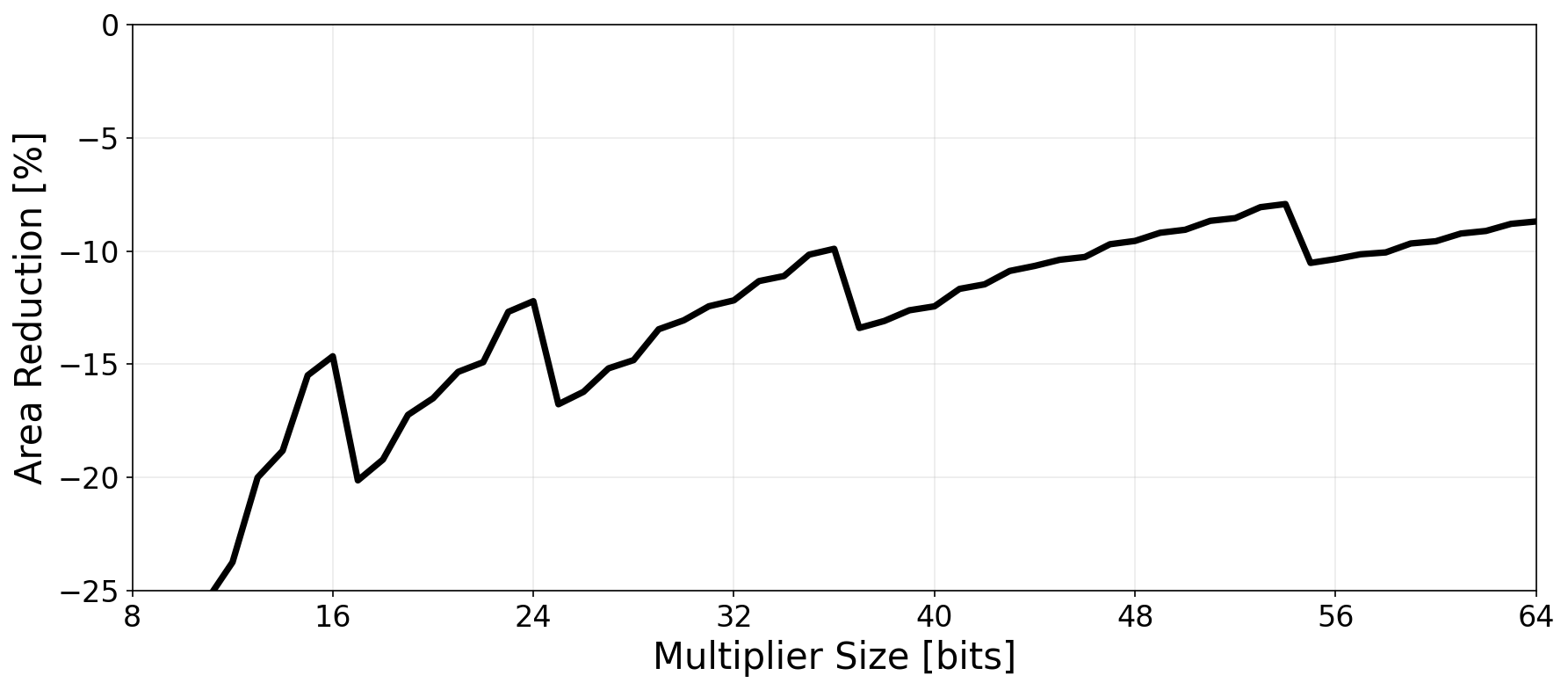}}
\caption{Dadda tree estimated logic volume reduction}
\label{fig06}
\end{figure}

Another aspect of this generator is area estimation. We can assume the area of the CSA3:2 being twice of a CSA2:2 and give an estimated reduction benefit as a function of multiplier size as shown in Fig.~\ref{fig06}. The estimated logic-volume reduction ranges from approximately 8\% to 25\% over the plotted multiplier sizes. Even though this estimate shows significant improvement, after a production EDA flow we typically see far fewer benefits \cite{burgess12}. Compression trees are very large cones of logic and pushing LE earlier in a cone of logic is known to improve the Electrical Effort (EE), or gain, along the cone \cite{le91}. Since a Wallace tree compresses as much as possible at every stage, EDA tools take advantage of this by building more gain earlier along the paths. As a result, the later stages can afford more complex standard cells with poor gain and skew the logical pressure per cell area metric in its favor. This phenomenon recovers a few percents but cannot beat having significantly less logic overall as shown in Fig.~\ref{fig06}.

\section{Beyond The Lone Multiplication} \label{sec:Beyond}

The real advantage of this methodology comes next. After all, modern
synthesis tools are very good at choosing the best standalone integer
multiplier layout and topology for a given micro-architecture
context \cite{miyasaka25}. But the reality is that high performance general
purpose vector units or specialized matrix accelerators will not have
``A * B'' directly written in the favored Hardware Design Language (HDL)
of the project. Redundant formats may be shared, accumulation vectors
may come later -or earlier- in a cycle and may need to be compressed
with other RDFs, more than one multiplication may happen at the same
time and they may need to be compressed together before the CPA, and so
on. There are so many use cases, we do not know, but one thing is for
sure: Compression trees and CPAs are functionally decoupled in practice.
In an effort to capture this compression tree microarchitecture
complexity, as an example, we will use a typical integer dot
product operator to implement as expressed in~\eqref{eq4}, with A and B being
vectors of D elements.

\begin{equation}
R = \sum_{d=0}^{D-1} A_d \cdot B_d\label{eq4}
\end{equation}

\begin{figure}[htbp]
\centerline{\includegraphics[width=0.8\linewidth]{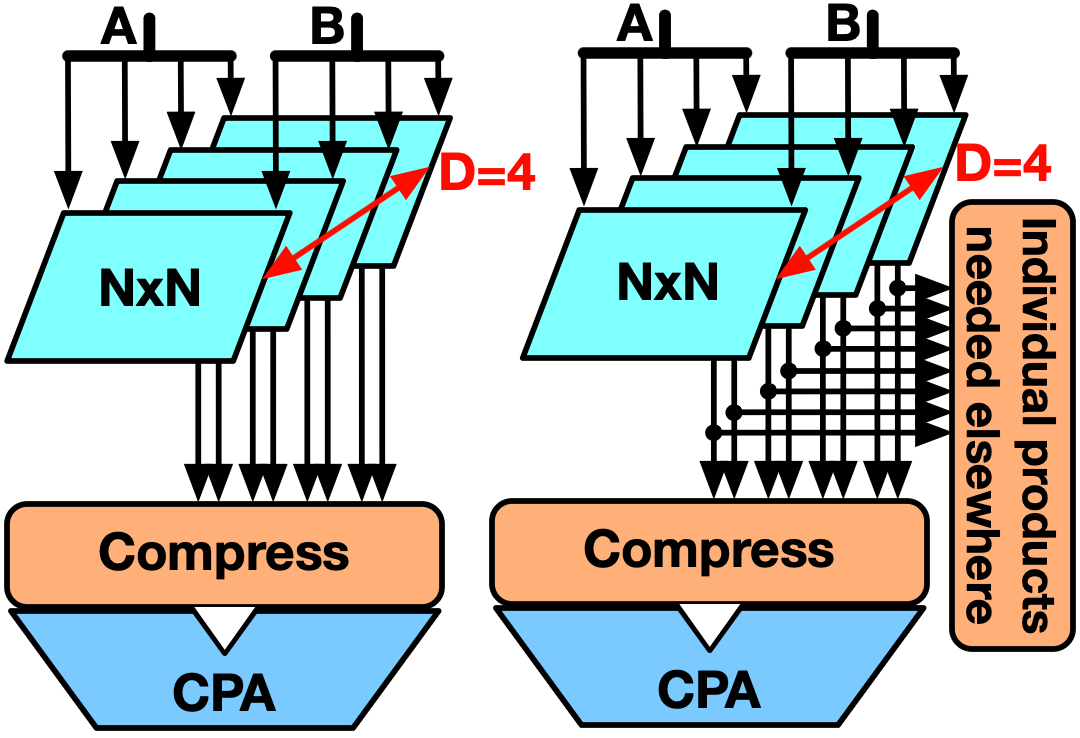}}
\centerline{(a)\hspace{80pt}(b)\hspace{20pt}}
\caption{Hypothetical micro-architectural setups for a 4-way dot product with NxN multipliers}
\label{fig07}
\end{figure}

To illustrate the situations compression trees may be confronted to, we
show two implementations of the operation in~\eqref{eq4} as shown in Fig.~\ref{fig07} for the case of D = 4:
\begin{itemize}
\item Straightforward fully fused summation of products.
\item The individual RDFs of each product are needed and shared with some other hypothetical datapaths in a way that is assumed to be less timing critical for the sake of this exercise.
\end{itemize}

In both cases of Fig.~\ref{fig07}, a straightforward RTL description of~\eqref{eq4} may cause synthesis to materialize each product through an intermediate CPA before performing the final summation, depending on the tool, library, and coding style. It would lead to slower and bigger designs with unnecessary CPAs. This is a classic case of forced decoupling of the compression tree and the CPA, and so, we may be left with the task of correctly generating the tree(s).

\begin{figure}[htbp]
\centerline{\includegraphics[width=1\linewidth]{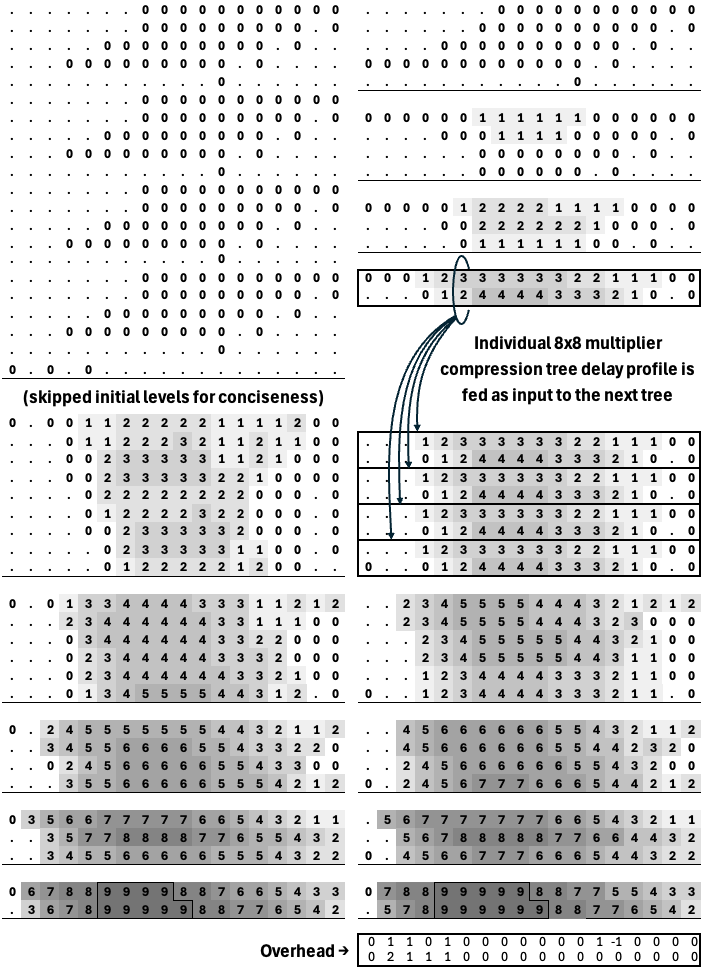}}
\centerline{(a)\hspace{120pt}(b)}
\caption{Comparative analysis of tree generation methodologies for two similar dot product summation micro-architectures}
\label{fig08}
\end{figure}

With this generator method, we can simply specify the matrix of relevant bits.
For the example of Fig.~\ref{fig07}a, we can write out all the partial products of all four multiplications together as shown in the top of Fig.~\ref{fig08}a and then let the generator sort out the best known arrangements of bits for each column based on our previously described methodology. Unfortunately, the bits are all scrambled and there is no practical way to extract the individual products even in redundant form.

Fig.~\ref{fig07}b example captures this micro-architectural constraint. For this case we must break apart the tree and first generate the small multipliers as shown in the top part of Fig.~\ref{fig08}b. But then we have the delay profile of the individual products RDFs, and we can use this information as the starting point for the next tree. That way, the dependent 8:2 compression tree can be generated with a complete knowledge of the delay profile of all its inputs.

\begin{figure}[htbp]
\centerline{\includegraphics[width=1\linewidth]{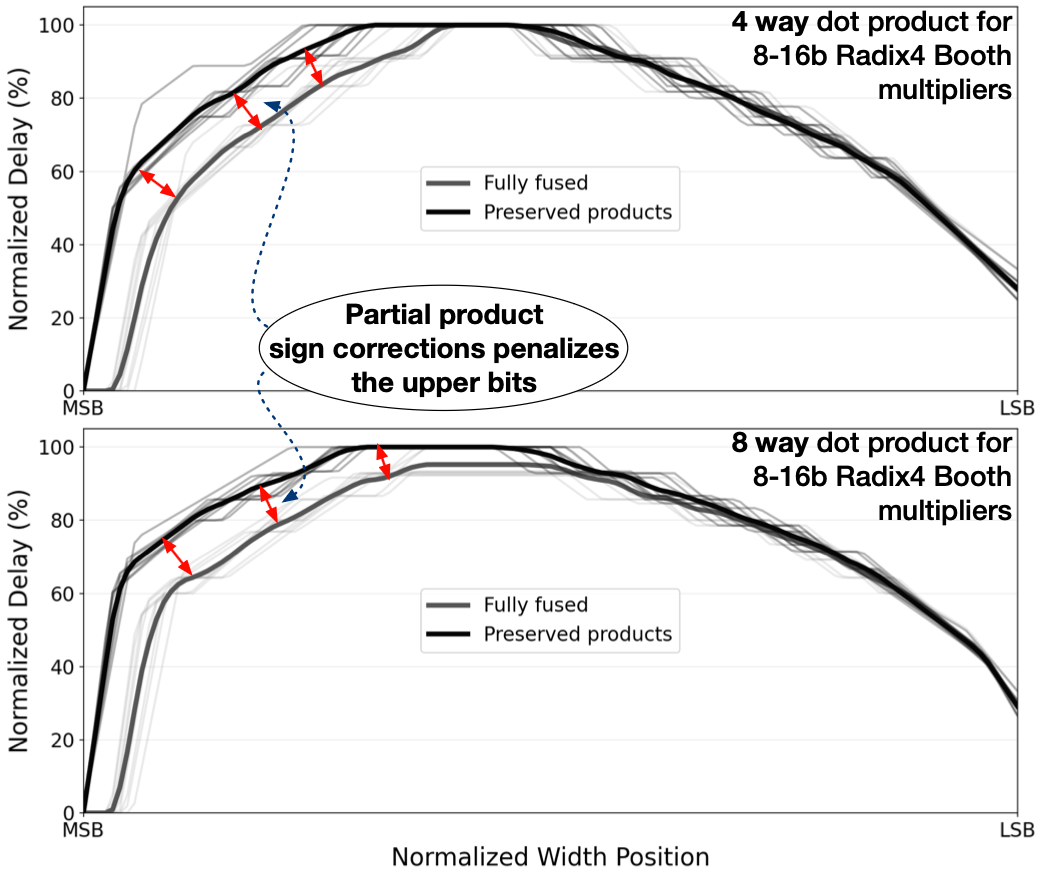}}
\caption{Relative delay results for fully fused vs. preserved products of 4-way and 8-way dot product trees with multipliers sizes varying from 8b to 16b}
\label{fig09}
\end{figure}

Unfortunately, the preserved products of Fig.~\ref{fig08}b must include the sign propagation logical ``1'' as part of the fully formed RDF for each individual product. These bits cannot be optimized away like the fully fused implementation in Fig.~\ref{fig08}a. Because of this limitation we can see a moderate delay overhead on the upper bits of the RDF result. 

If we extend this study further by varying the multipliers from 8b to 16b, still encoded as Radix-4 Booth, and also change the depth of the dot product to 8 ways, we can clearly see the trends in Fig.~\ref{fig09}: 
\begin{itemize}
\item The delay of the lower half of the tree is not affected by the constraint of extracting the individual multipliers RDF, regardless of multiplier size and dot product depth.
\item The overhead noticed in Fig.~\ref{fig08} for the case of \{D;N\}=\{4;8\} remains proportional and consistent across the tried D and N dimensions.
\item The overhead of these signed propagation bits starts to marginally affect the critical path when D=8 if we wish to preserve the intermediate products RDFs.
\end{itemize} 

\begin{figure}[htbp]
\centerline{\includegraphics[width=1\linewidth]{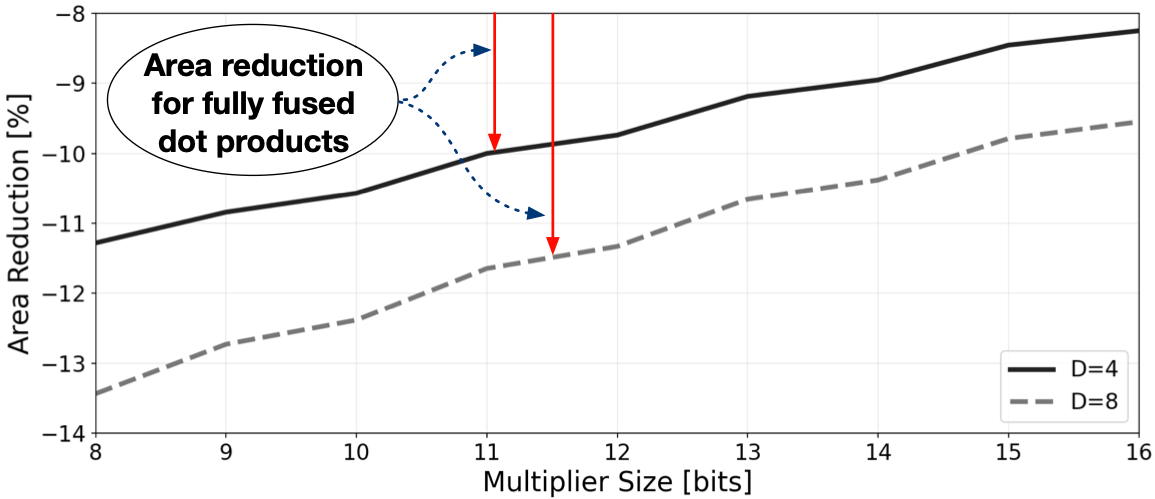}}
\caption{Estimated Area-reductions results for fully fused vs. preserved products of 4-way and 8-way dot product trees from 8b to 16b}
\label{fig10}
\end{figure}

For the area estimation, we can also see in Fig.~\ref{fig10} a trend consistent with Fig.~\ref{fig06}. As N grows the generated tree structure becomes wider and we see diminishing returns as the majority of the logic becomes less impacted by the overhead. However, as we increase the depth of the dot product from D=4 to 8, the overhead compounds linearly which is why we see a proportional area reduction offset across all multiplier sizes. The same compounded overhead effect is also the reason the critical path can eventually be affected for higher dot product depth.

\section{The Late Accumulate Case} \label{sec:LateAcc}

Another important ``beyond multiplication'' common case is the late accumulation problem statement: if we were to accumulate one vector with the multiplier RDF before the final CPA, how should we proceed?
If the vector is early or no later than the partial products from the Booth encoders, then it should be included as one of the vectors fed to the compression tree at time 0. Of course, if it arrives later than the multiplier output RDF delay profile, then the accumulator is the critical path and the compression tree timing is irrelevant.

But for the in-between cases, when the accumulator timing profile coincides somewhere during the compression tree, we need to understand the policy: Should we annotate that ``later than time 0'' vector and let the Dadda tree reduce, break up the tree and insert it somewhere in the middle, or simply add it at the end anyway with a row of CSA3:2? It turns out, parallel compression trees have the capacity of absorbing some of that delay until it becomes worse than doing that row of CSA3:2 at the end. A late arrival threshold exists where, for a particular Eq.~\ref{eq3} reduction stage, the late vector is forcefully added in that stage even though the critical path of the multiplier delay profile is significantly lower. This pushes both sums and carries of the next stage to be slower, effectively propagating the slowest path at a 2x rate throughout all the branches of the tree.

\begin{figure}[htbp]
\centerline{\includegraphics[width=1\linewidth]{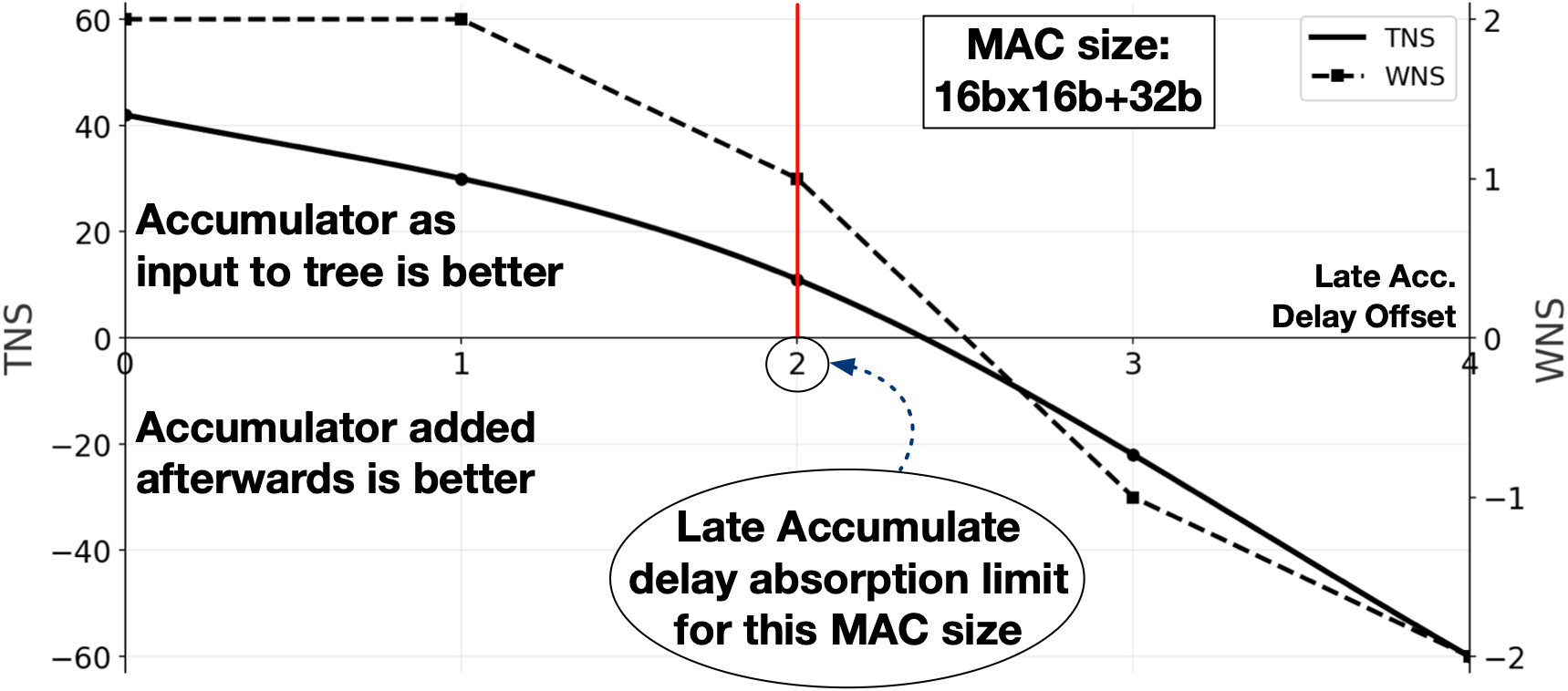}}
\caption{Late accumulate delay absorption of a 16x16 multiply accumulate}
\label{fig11}
\end{figure}

Fundamentally, these compression trees are meant to be parallel balanced structures... So bringing in a noticeably late input, i.e. more than the maximum 2 delay units of a stage, challenges the base parallelism precepts.
Nevertheless, with this methodology, we can estimate the late accumulate absorption capacity of a compression tree as a function of the multiplier size.
For example, in the case of a 16x16 with a 32b accumulator, we can compare the delay outcome when the accumulator is an input to the tree just like the partial products versus adding it at the end with a row of CSA3:2. Fig.~\ref{fig11} shows the absorption limit as, expectedly, a maximum of 2 delay units for the 16x16 case. In this comparison, ``TNS'' represents the cumulative difference of arrival times and ``WNS'' the largest difference, as opposed to their traditional meaning with reference to the clock. 

\begin{figure}[htbp]
\centerline{\includegraphics[width=1\linewidth]{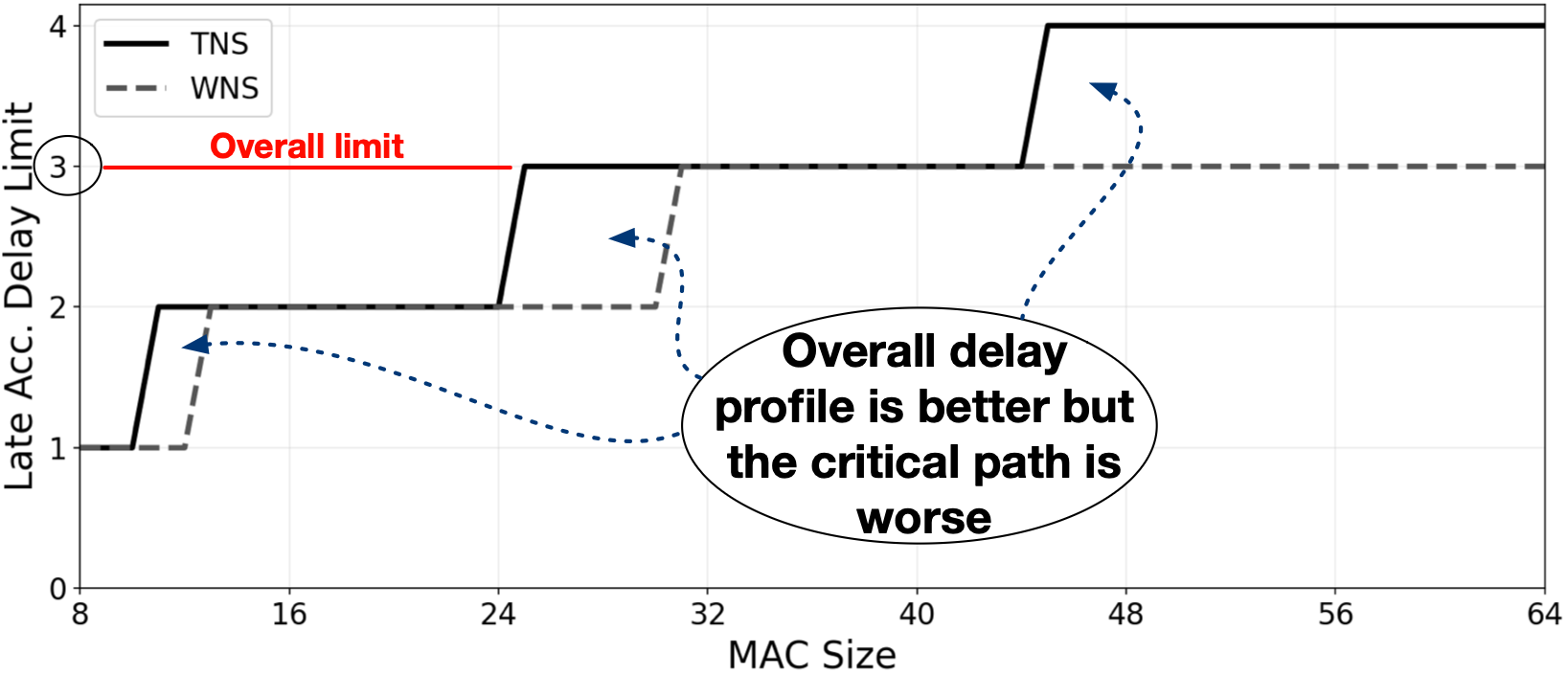}}
\caption{Late accumulate delay absorption limits across all MAC sizes}
\label{fig12}
\end{figure}

We can push this analysis even further, and see across all multiplier sizes how that limit changes. Fig.~\ref{fig12} proves a limited availability in absorption: Even for large multipliers, the maximum late arrival of an accumulator maxes out at 3 units for the critical path. Interestingly, in some cases, even though the critical path ends up being worse, the output delay profile with the accumulator as input can still have a positive cumulative difference compared to accumulating afterwards. Either way, very quickly a slow input has a very negative impact to the overall performance of the tree. Of course, as the multiplier gets smaller, the tree is less accommodating to late arrivals. Perhaps, it would be preferable to estimate the appropriate insertion point in the middle of the tree for late arrivals. That way, the compression tree can break the Eq.~\ref{eq3} iteration in order to plug in late vector(s) efficiently.

\section{Next Step}
Using Delay as the sole dimension for the weights of a bit heap can still be sub-optimal. A second dimension of weights can be tracked alongside the delay: Switching activity.

Two-dimensional optimization of Delay reduction and Energy usage becomes a more complex topic since the answer is a design space as opposed to a particular design configuration \cite{giacomotto07}. Nevertheless, we can split the problem in 2 steps:

\begin{enumerate}
\def\labelenumi{\arabic{enumi})}
\item
Make the switching activity strictly a secondary grouping criterion: We may still have the opportunity to postpone/reorder activity nodes based on their activity weights at no timing cost by only manipulating the already selected delay equivalent sub-groups from the existing delay-aware heuristic.
\item
Derate delay, by increasing logical depth, on some paths to improve overall switching activity. However, higher logical depth typically contribute to higher energy usage from higher logical effort which is counterproductive to the overall switching activity reduction effort. In this case, we start to trade delay for energy and we fall into the design space solution which involves technology nodes and physical design awareness. Furthermore, if the tree has plenty of room in delay to be primarily activity-aware constructed, maybe the external micro-architecture should be revisited or repipelined. Conveniently, this effort would bring us back to case 1).
\end{enumerate}


\section{Conclusion}
Compression trees should be considered general-purpose bit-heap reduction structures rather than components that are exclusively associated with multipliers. As arithmetic datapaths increasingly combine multiple products, accumulations, and redundant representations, the ability to generate a compression tree independently of the surrounding arithmetic operation becomes increasingly important.

This work presented a delay-aware methodology based on weighted bit heaps, where each bit is characterized by its position and relative arrival time. Using the flexibility of the basic CSA3:2 operator, the proposed approach preserves the maximum practical bandwidth between reduction stages, following the fundamental principle of the Dadda topology. This organization allows early-arriving bits in less-populated columns to be used opportunistically in later stages, reducing logical pressure on critical paths. For the evaluated Radix-4 Booth multipliers, this approach consistently produced lower delay and reduced estimated logic volume than other strategies. It also demonstrated that a Dadda-style tree can absorb additional bits or uneven arrival times with less penalty on the critical path.
The same methodology applies beyond standalone multiplication. By describing only the relevant weighted bits, the generator can evaluate fused dot products, shared redundant representations, and other datapaths in which compression trees and carry-propagating adders are naturally decoupled. It can also quantify the cost of microarchitectural constraints, such as requiring individual intermediate products to remain available. This provides a useful way to explore arithmetic organizations before committing to a particular physical implementation design flow.

The delay model used here is intentionally technology-independent and should therefore be regarded as a logical-pressure metric rather than a replacement for detailed physical implementation analysis. Nevertheless, it provides a simple and versatile basis for comparing candidate trees and guiding production datapath generation. Future work will extend the weighted bit-heap model with switching activity, enabling energy-aware optimization while retaining delay as the primary criterion. Such an extension could support explicit tradeoffs among timing, area, and energy, as well as provide a more complete framework for optimizing compression trees in the context of modern vector and matrix-processing datapaths.
~\\
~\\

\printbibliography

\end{document}